\documentclass[fleqn,10pt]{wlscirep}
\title{The sinking of the El Faro: predicting real world rogue waves during Hurricane Joaquin}

\author[1,*]{Francesco Fedele}
\author[2,3]{Claudio Lugni}
\author[4]{Arun Chawla}
\affil[1]{School of Civil \& Environmental Engineering, Georgia Institute of Technology, Atlanta, Georgia 30332, USA}
\affil[2]{CNR-INSEAN \& Marine Technology Center - Italian Research Council, Roma 00128, Italy}
\affil[3]{NTNU-AMOS \& Center for Autonomous Marine Operation Systems, Trondheim 7491, Norway}
\affil[4]{National Center for Weather \& Climate Prediction, Marine Modelling \& Analysis Branch, College Park 20740, USA}
\affil[*]{Corresponding author's email: fedele@gatech.edu}

\begin{abstract}
\emph{We present a study on the prediction of rogue waves during the 1-hour sea state of Hurricane Joaquin when the Merchant Vessel El Faro sank east of the Bahamas on October 1, 2015. High-resolution hindcast of hurricane-generated sea states and wave simulations are combined with novel probabilistic models to quantify the likelihood of rogue wave conditions. 
The data suggests that the El Faro vessel was drifting at an average speed of approximately~$2.5$~m/s prior to its sinking. As a result, we estimated that the probability that El Faro encounters a rogue wave whose crest height exceeds 14 meters while drifting over a time interval of 10~(50) minutes is $\sim1/400$~$(1/130)$. 
The largest simulated rogue wave has similar generating mechanism and characteristics of the Andrea, Draupner and Killard rogue waves as the constructive interference of elementary waves enhanced by bound nonlinearities.}
\end{abstract}
\begin{document}

\flushbottom
\maketitle
%
%
\thispagestyle{empty}


\section*{Introduction}

The tragic sinking of the SS El Faro vessel occurred while it was traveling from Florida to Puerto Rico~\cite{elfaro}. 
The vessel with a crew of 33 sank about 1140 Hrs UTC on Oct. 1, 2015. As part of their investigation into the sinking of the El Faro, the National Transportation Safety Board (NTSB) has requested us an analysis on the occurrence of rogue waves during Hurricane Joaquin around the time and location of the El Faro's sinking~\cite{fedele2016prediction}. Here, we present the main results of our rogue wave analysis. 

The data suggests that the El Faro vessel was drifting at an average speed of approximately~$2.5$~m/s prior to its sinking~\cite{fedele2016prediction}. As a result, El Faro has a higher probability to encounter a rogue wave while drifting over a period of time than that associated with a fixed observer at a point of the ocean. Indeed, the encounter of a rogue wave by a moving vessel is analogous to that of a big wave that a surfer is in search of~\cite{Fedele2012,fedele2013}. The surfer's likelihood to encounter a big wave increases if he moves around a large area instead of staying still. Indeed, if he spans a large area the chances to encounter a large wave increase. This is a space-time effect very important for ship navigation and it cannot be neglected. Such an effect is considered in our rogue wave analysis by way of a new probabilistic model for the prediction of rogue waves encountered by a vessel along its navigation path~\cite{Fedele2012,fedele2015JPO}. In particular, we give a theoretical formulation and interpretation of the exceedance probability, or occurrence frequency of a rogue wave by a moving vessel. 

\begin{table}[ht]
\begin{centering}
\resizebox{\textwidth}{!}{
\begin{tabular}{lllll}
\hline 
 & \textbf{El Faro} & \textbf{Andrea} & \textbf{Draupner} & \textbf{Killard}\tabularnewline
\hline 
\hline 
Significant wave height $H_{s}$~{[}m{]} & 9.0 & 10.0 & 11.2 & 11.4\tabularnewline
\hline 
Dominant wave period $T_{p}$~{[}s{]} & 10.2 & 14.3 & 15.0 & 17.2\tabularnewline
\hline 
Mean zero-crossing wave period $T_{0}$~{[}s{]} & 9.2 & 11.1 & 11.3 & 13.2\tabularnewline
\hline 
Mean wavelength $L_{0}$ {[}m{]} & 131 & 190 & 195 & 246\tabularnewline
\hline 
Depth $d$~{[}m{]}, $k_0 d$ with $k_0=2\pi/L_0$ & 4700, 2.63 & 74, 2.23 & 70, 2.01 & 58, 1.36\tabularnewline
\hline 
Spectral bandwidth $\nu$ & 0.49 & 0.35 & 0.36 & 0.37 \tabularnewline
\hline 
Angular spreading $\theta_{\nu}$ & 0.79 & 0.43 & 0.44 & 0.39\tabularnewline
\hline 
Parameter~$R=\theta_{\nu}^{2}/2\nu^{2}$ ~\cite{Janssen2009} & 1.34 & 0.72 & 0.75 & 0.56\tabularnewline
\hline 
Benjamin Feir Index $BFI$ in deep water ~\cite{Janssen2003} & 0.36 & 0.24 & 0.23 & 0.18\tabularnewline
\hline 
Tayfun NB skewness~$\lambda_{3,NB}$~\cite{Tayfun2006} & 0.26  & 0.159 & 0.165 & 0.145\tabularnewline
\hline 
Mean skewness $\lambda_{3}$ from HOS simulations & 0.162 & 0.141 & 0.146 & 0.142\tabularnewline
\hline 
Maximum NB dynamic excess kurtosis 
$\lambda_{40,\textit{max}}^d$
~\cite{fedele2015kurtosis} &$ 10^{-3}$ & $1.3\cdot10^{-3}$ & $1.1\cdot10^{-3}$ & $1.6\cdot10^{-3}$\tabularnewline
\hline
Janssen NB bound excess kurtosis~$\lambda_{40,NB}^{d}$ ~\cite{JanssenJFM2009} & 0.049 & 0.065 & 0.074 & 0.076\tabularnewline
\hline
Mean excess kurtosis $\lambda_{40}$ from HOS simulations & 0.042 & 0.041 & 0.032 & $-0.011$ \tabularnewline
\hline 
\hline 

Actual maximum crest height $h/H_{s}$& 1. 68  & 1.55 & 1.63 & 1.62\tabularnewline
\hline 
Actual maximum crest-to-trough (wave) height $H/H_{s}$ & 2.6 & 2.30 & 2.15 & 2.25\tabularnewline
\hline
\hline 
\end{tabular}
}
\par\end{centering}

\protect\caption{Wave parameters and various statistics of the simulated El Faro sea state in comparison to the Andrea, Draupner and Killard rogue sea states~\cite{FedeleSP2016}. We refer to the Methods section for the definitions of the wave parameters.}

\end{table}

\begin{figure}[h]
\centering\includegraphics[scale=0.5]{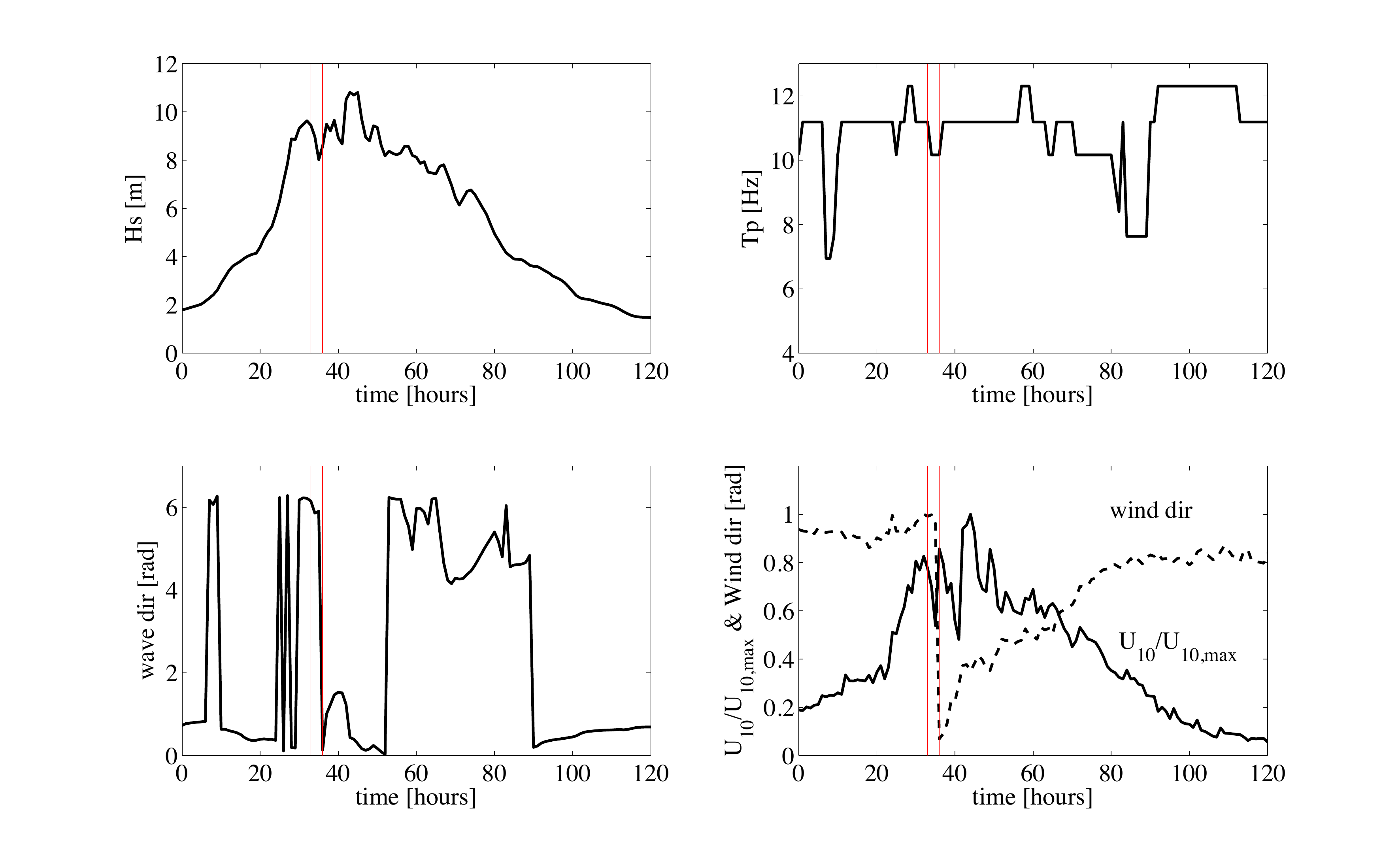} 
\caption{WAVEWATCH III parameters history during Hurricane Joaquin around the location where the El Faro vessel sank. (top-left) Hourly variation of the significant wave height $H_s$, (top-right) dominant wave period $T_p$, (bottom-left) dominant wave direction and (bottom-right) normalized $U_{10}/U_{10,max}$ wind speed (solid line) and direction (dashed line). Maximum wind speed $U_{10,max}=51 m/s$. Red vertical lines delimit the 1--hour interval during which the El Faro vessel sank.} 
\label{FIG1}
\end{figure}


\begin{figure}[h]
\centering\includegraphics[scale=0.5]{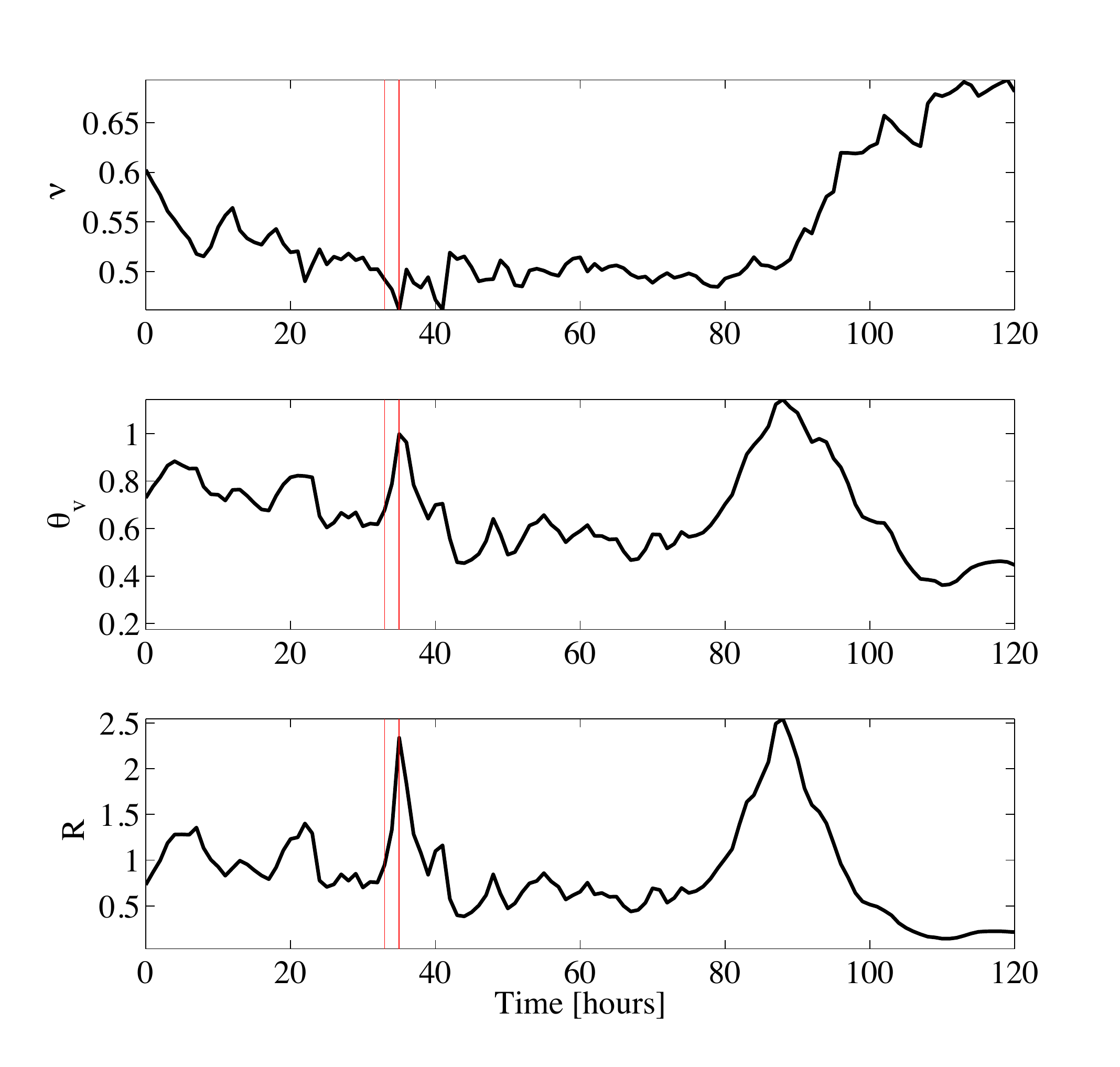} 
\caption{WAVEWATCH III parameters history during Hurricane Joaquin around the location where the El Faro vessel sank. (top) Hourly variation of the spectral bandwidth $\nu$ history, (center) directional spreading $\theta_v$ and (bottom) directional factor $R=\frac{1}{2}\theta_v^2/\nu^2$. Red vertical lines delimit the 1-hour interval during which the El Faro vessel sank.}
\label{FIG3}
\end{figure}

\begin{figure}[h]
\centering\includegraphics[scale=0.5]{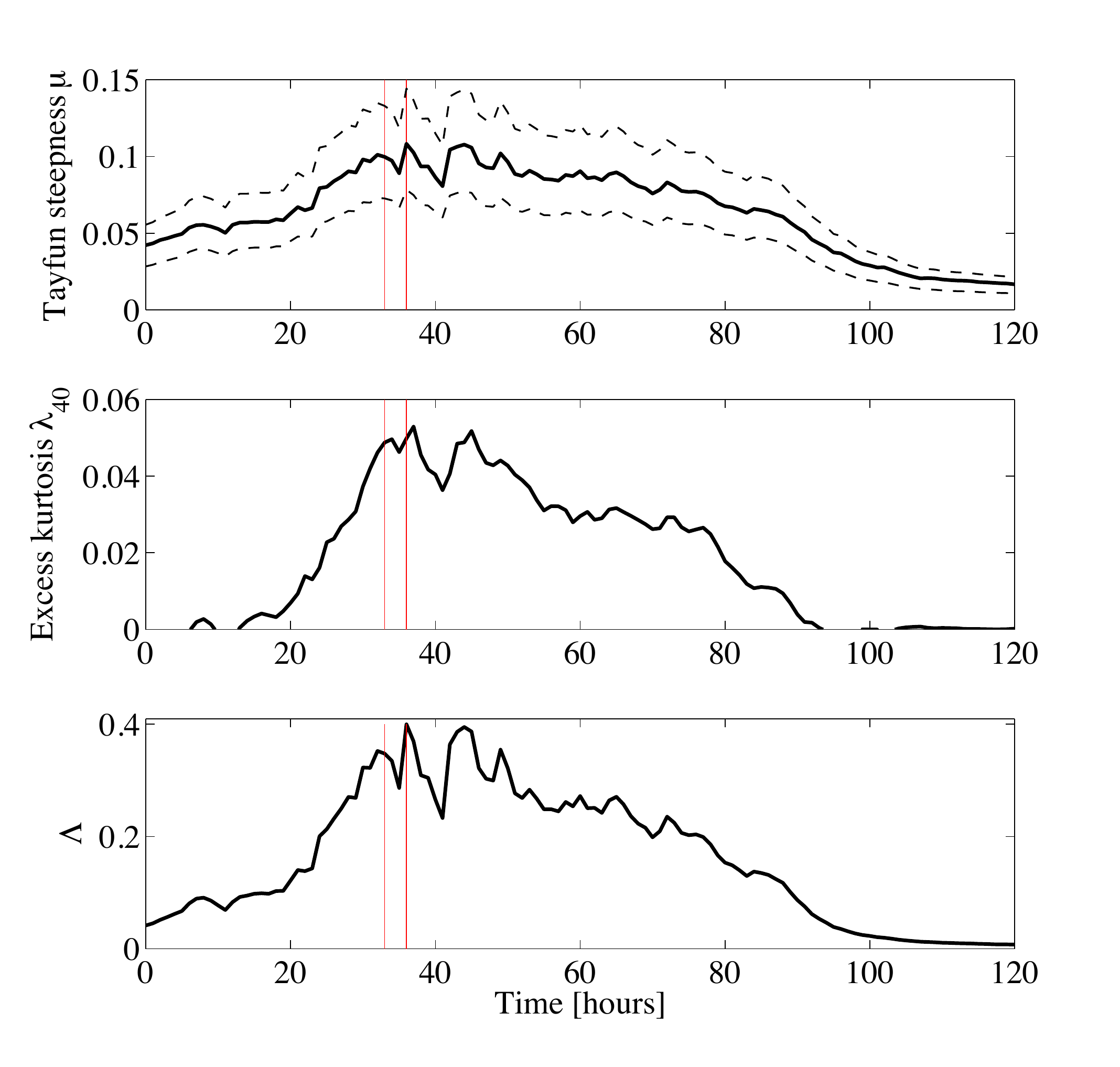} 
\caption{WAVEWATCH III parameters history during Hurricane Joaquin around the location where the El Faro vessel sank. (top) Hourly variation of the Tayfun steepness $\mu$ (solid line) with bounds (dashed lines), (center) excess kurtosis $\lambda_{40}$ and (bottom) nonlinear coefficient $\Lambda\sim 8\lambda_{40}/3$. Red vertical lines delimit the 1-hour interval during which the El Faro vessel sank.}
\label{FIG4}
\end{figure}

%

\section*{Results}

Our rogue wave analysis focused on the study of the 1-hour sea state of Hurricane Joaquin during which the El Faro vessel sank, hereafter referred to as the El Faro sea state. The convenient wave parameters and statistical models are defined in the Methods section.

\subsection*{Metaocean parameters of Hurricane Joaquin in the region of the sinking of El Faro}

We use the hindcast directional spectra by WAVEWATCH III and describe the wave characteristics of the sea states generated by Hurricane Joaquin about the time and location where the El Faro vessel sank~\cite{NTSB_meteo}. 
The top panel on the left of Fig.~(\ref{FIG1}) shows  hourly variation of the significant wave height $H_s$ during the event. The top-right panel displays the history of the dominant wave period $T_p$, and the dominant wave direction, the neutral stability 10-m wind speed $U_{10}$ and direction are shown in the bottom-panels respectively. The red vertical lines delimit the 1--hour interval during which the El Faro vessel sank. 

The encountered 1-hour sea state by El Faro about the time and location of sinking had a significant wave height of $H_s\approx9$~m and the maximum wind speed was $U_{10,max}=51$~m/s. It was very multidirectional (short-crested) as indicated by the large values of both the spectral bandwidth $\nu$ and angular spreading $\theta_v$ as shown in~Fig.~(\ref{FIG3}). 

In Table 1 we report the metaocean parameters of the El Faro sea state in comparison to those of the Draupner, Andrea and Killard rogue sea states~\cite{FedeleSP2016}.  Note that the four sea states have similar metaocean characteristics. However, El Faro is a steeper sea state as the mean wavelengh $L_0$ is shorter than the other three states. 

\subsection*{Statistical properties of Hurricane Joaquin-generated seas}\label{STAT}

The relative importance of ocean nonlinearities can be measured by integral statistics such as the wave skewness $\lambda_3$ and the excess kurtosis $\lambda_{40}$ of the zero-mean surface elevation $\eta(t)$. The skewness describes the effects of second-order bound nonlinearities on the geometry and statistics of the sea surface with higher sharper crests and shallower more rounded troughs~\cite{Tayfun1980,TayfunFedele2007,Fedele2009}. The excess kurtosis comprises a dynamic component $\lambda_{40}^{d}$ measuring third-order quasi-resonant wave-wave interactions and a bound contribution $\lambda_{40}^{b}$ induced by both second- and third-order bound nonlinearities~\cite{Tayfun1980,TayfunLo1990,TayfunFedele2007,Fedele2009,Fedele2008,Janssen2009}.

In deep waters, the dynamic kurtosis~\cite{fedele2015kurtosis} depends on the Benjamin-Feir index $BFI$ and the parameter $R$, which is a dimensionless measure of the multidirectionality of dominant waves~\cite{Janssen2009,Mori2011,fedele2015kurtosis}. For unidirectional (1D) waves $R=0$. 
The bottom panel of Fig.~(\ref{FIG3}) displays the hourly variations of the directional factor $R$ during Hurricane Joaquin near the location where El Faro sank. About the peak of the hurricane the generated sea states are very multidirectional (short-crested)  as $R>1$ and so wave energy can spread directionally. As a result, nonlinear focusing due to modulational instability effects diminishes~\cite{fedele2015kurtosis,Onorato2009,WasedaJPO2009,Toffoli2010} and becomes essentially insignificant under such realistic oceanic conditions~\cite{Shrira2013_JFM,Shrira2014_JPO,fedele2015kurtosis,FedeleSP2016}.  

The top panel of Fig.~(\ref{FIG4}) displays the hourly variation of the Tayfun steepness $\mu$ (solid line) with bounds (dashed lines). The excess kurtosis $\lambda_{40}$ mostly due to bound nonlinearities is shown in the center panel and the associated $\varLambda$ parameter at the bottom. The red vertical lines delimit the 1-hour interval during which the El Faro vessel sank.

In Table 1 we compare the statistical parameters of the El Faro sea state and the Draupner, Andrea and Killard rogue sea states~(from~\cite{FedeleSP2016}). Note that the El Faro sea state has the largest directional spreading. Moreover, for all the four sea states the associated $BFI$ are less than unity and the maximum dynamic excess kurtosis is of~$O(10^{-3})$ and thus negligible in comparison to the associated bound component. Thus, third-order quasi-resonant interactions, including NLS-type modulational instabilities play an insignificant role in the formation of large waves~\cite{fedele2015kurtosis, FedeleSP2016} especially as the wave spectrum broadens~\cite{Fedele2014} in agreement with oceanic observations available so far~\cite{TayfunFedele2007,Tayfun2008,Christou2014}. On the contrary, NLS instabilities have been proven to be effective in the generation of optical rogue waves~\cite{Dudley2016}. 


\subsection*{Higher Order Spectral (HOS) simulations of the El Faro sea state}
\begin{figure}[h]
\centering\includegraphics[scale=0.42]{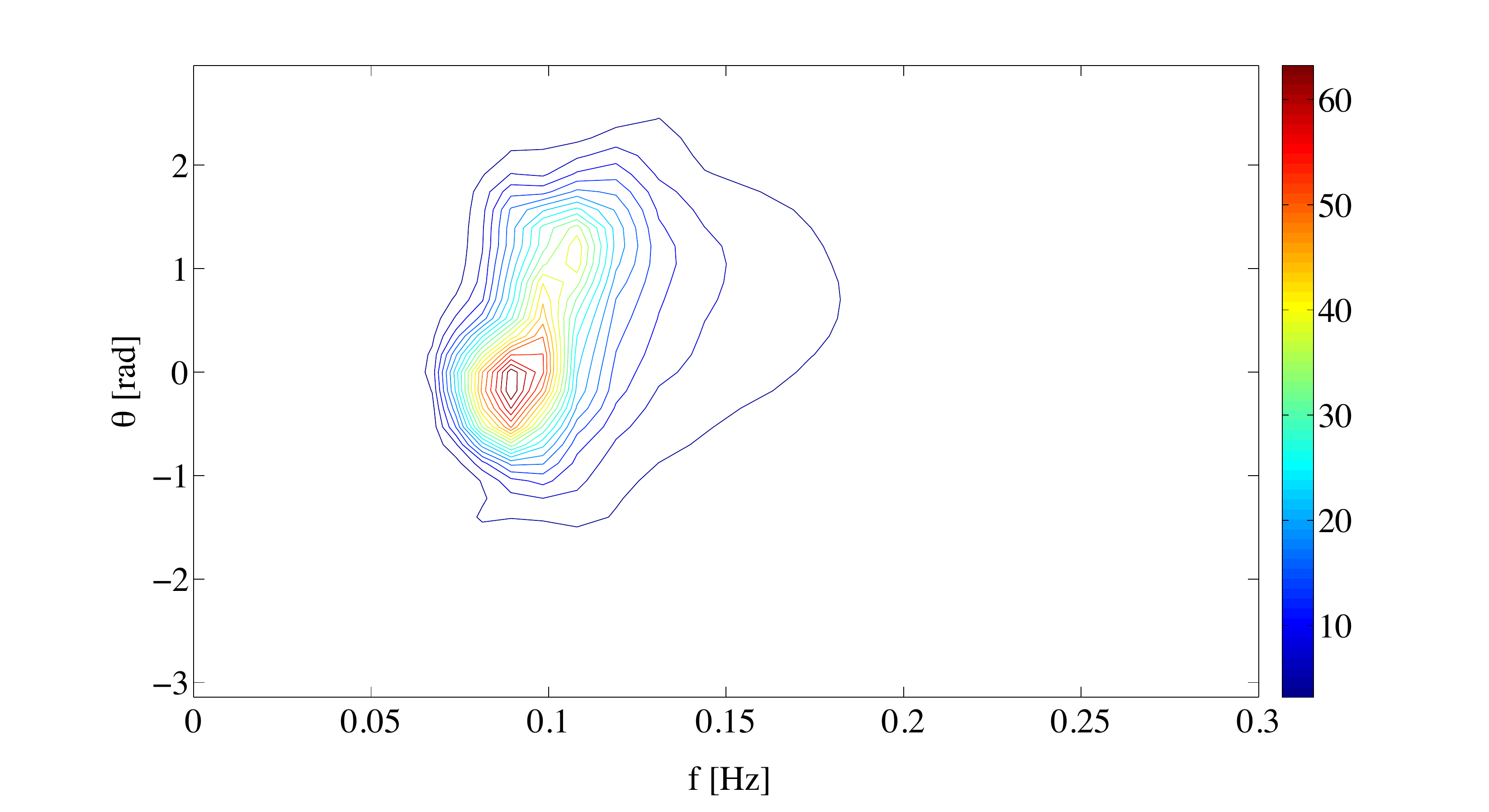} 
\caption{WAVEWATCH III hindcast directional spectrum ${S}(f,\theta)$~$[m^2 s/rad]$ at approximately the time and location of the El-Faro sinking.} 
\label{FIG_WWIII}
\end{figure}

We have performed Higher-Order pseudo-Spectral (HOS) simulations~\cite{DommermuthYue1987HOS,West1987} of the El Faro sea state over an area of~$4$~km~x~$4$~km for a duration of 1 hour (see Methods section for a description of the numerical method). The initial wave field conditions  are defined by the WAVEWATCH III hindcast directional spectrum ${S}(f,\theta)$ around the time and region of the El Faro sinking as shown in Fig.~\ref{FIG_WWIII}. Our HOS simulations are performed accounting only for the full (resonant and bound) nonlinearities of the Euler equations up to fourth order in wave steepness.

 
\begin{figure}[h]
\centering\includegraphics[scale=0.6]{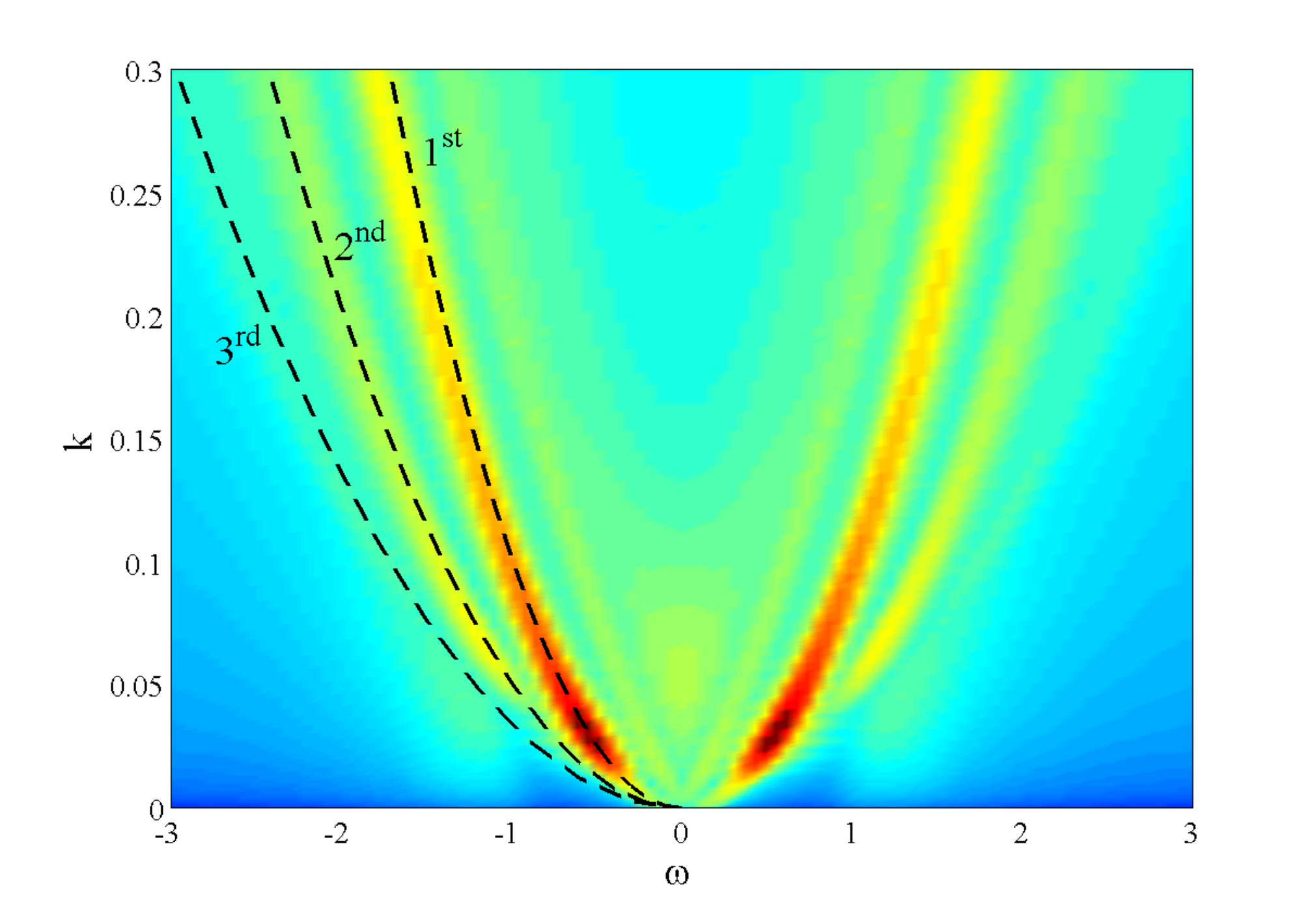} 
\caption{HOS simulations of the El Faro sea state: predicted wavenumber-frequency spectrum~$S(k,\omega)$~$[m^2 s/rad]$. Sea state duration of 1 hour over an area of~$4$~km~x~$4$~km; the wave field is resolved using $1024$~x~$1024$ Fourier modes.
} 
\label{FIG_EKW}
\end{figure}

The wavenumber-frequency spectrum $S(k,\omega)$ estimated from the HOS simulations is shown in Figure~\ref{FIG_EKW}. 
Here, dashed lines indicate the theoretical dispersion curves related to the first-order ($1^{st})$ free waves as well as the second ($2^{nd})$ and third-order ($3^{rd}$) bound harmonic waves. The HOS predictions indicate that second order nonlinearities are dominant with a weak effect of third-order nonlinear bound interactions, in agreement with recent studies of rogue sea states~\cite{FedeleSP2016}. Further, fourth-order effects are insignificant.

The wave skewness and kurtosis rapidly reach a steady state in few wave mean periods as an indication that third-order quasi-resonant wave-wave interactions are negligible in agreement with theoretical predictions~\cite{fedele2015kurtosis} and simulations~\cite{FedeleSP2016}. Note that the theoretical narrowband predictions slightly overestimate the simulated values for skewness and excess kurtosis~(see Table 1). The same trend is also observed in recent studies on rogue waves~\cite{FedeleSP2016}. This is simply because narrowband approximations do not account for the directionality and finite spectral bandwidth of the El Faro wave spectrum.


\begin{figure}[h]
\centering\includegraphics[scale=0.75]{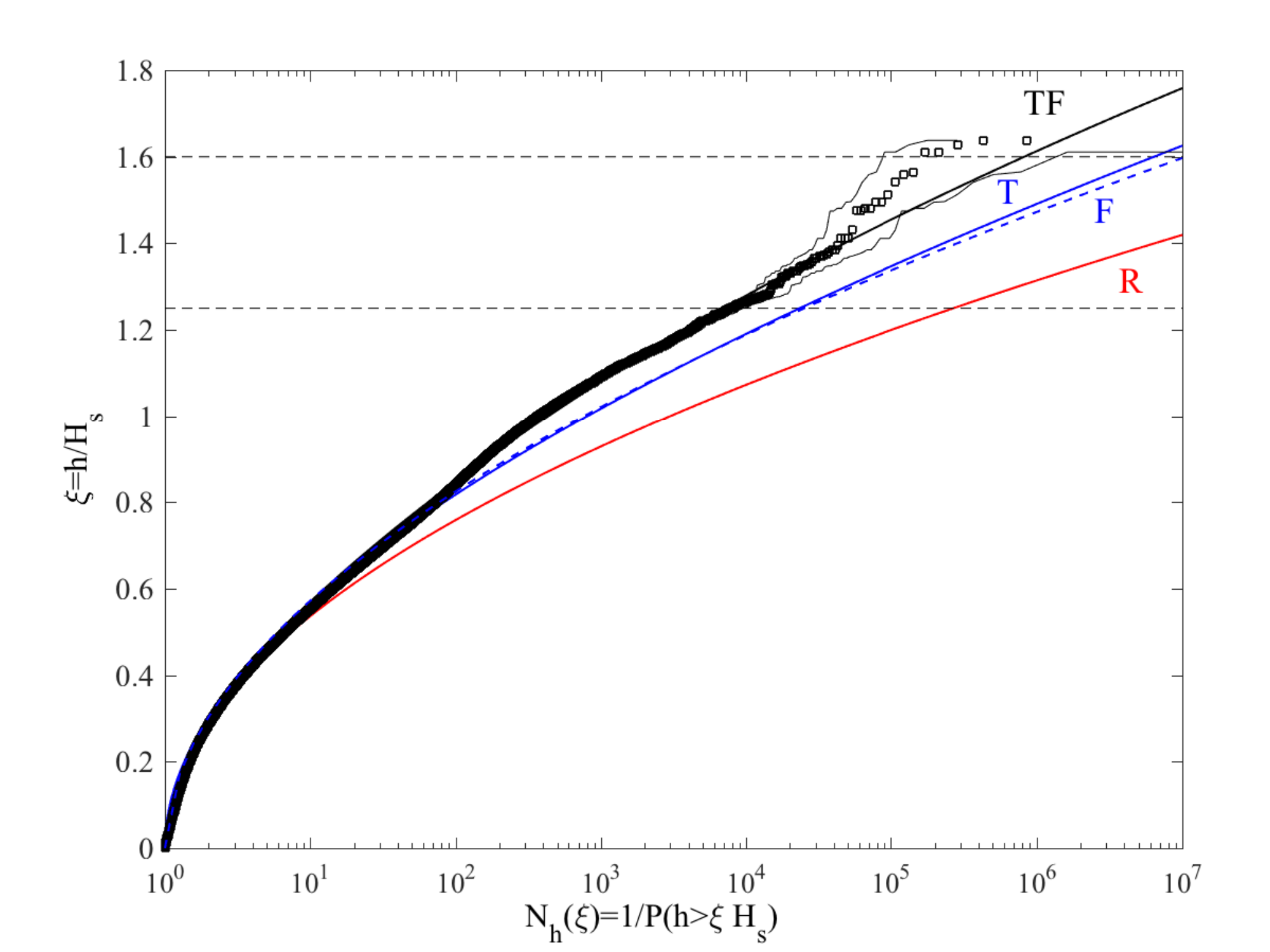} 
\caption{HOS simulations of the El Faro sea state. Crest height scaled by the significant wave height ($\xi$) versus conditional return period ($N_h$) for the (left) Andrea, (center) Draupner and (right) Killard rogue sea states:  HOS numerical predictions ($\square$) in comparison with theoretical models:F=Forristall (blue dashed) T=second-order Tayfun (blue solid), TF=third-order (red solid) and R=Rayleigh distributions (red solid). Confidence bands are also shown~(light dashes). $N_h(\xi)$ is the inverse of the exceedance probability $P(\xi)=\mathrm{Pr}[h>\xi H_s]$. Horizontal lines denote the rogue threshold~$1.25H_s$~\cite{DystheKrogstad2008} and~$1.6H_s$.} 
\label{FIG7}
\end{figure}

\subsection*{Occurrence frequency of a rogue wave by a fixed observer: the return period of a wave whose crest height exceeds a given threshold}\label{TimeProb}

To describe the statistics of rogue waves encountered by a fixed observer at a given point of the ocean, we consider the conditional return period $N_h(\xi)$ of a wave whose crest height exceeds the threshold $h=\xi H_s$, namely
\begin{equation}
N_h(\xi)=\frac{1}{\mathrm{Pr}\left[h>\xi H_s\right]}=\frac{1}{P(\xi)},\label{Nh}
\end{equation}
where $P(\xi)$ is the probability or occurrence frequency of a wave crest height exceeding~$\xi H_s$ as encountered by a fixed observer. In other words, $P(\xi)$ is the probability to randomly pick from a time series observed at a fixed point of the ocean a wave crest that exceeds the threshold $\xi H_s$. Equation~\eqref{Nh} also implies that the threshold~$\xi H_s$, with $H_s=4\sigma$, is exceeded on average once every $N_{h}(\xi)$ waves. For weakly nonlinear random seas, the probability $P$ is hereafter described by the third-order Tayfun-Fedele~\cite{TayfunFedele2007} (TF), second-order Tayfun~\cite{Tayfun1980} (T), second-order Forristall~\cite{Forristall2000} (F) and the linear Rayleigh (R) distributions~(see Methods section). 
%
%

Our statistical analysis of HOS wave data suggest that second-order effects are the dominant factors in shaping the probability structure of the El Faro sea state with a minor contribution of excess kurtosis effects. Such dominance is seen in Fig.~\ref{FIG7}, where the HOS numerical predictions of the conditional return period $N_h(\xi)$ of a crest exceeding the threshold $\xi H_s$ are compared against the theoretical predictions based on the linear Rayleigh (R), second-order Tayfun (T) and third-order (TF) models from~Eq.~\eqref{Pid} (sampled population of $10^6$ crest heights). In particular, $N_h(\xi)$ follows from Eq.~\eqref{Nh} as the inverse $1/P(\xi)$ of the empirical probabilities of a crest height exceeding the threshold $\xi H_s$. An excellent agreement is observed between simulations and the third-order TF model up to crest amplitudes $h/H_s\sim1.5$. For larger amplitudes, the associated confidence bands of the estimated empirical probabilities widen, but TF is still within the bands. Donelan and Magnusson~\cite{donelan2017} suggest that the TF model agrees with the Andrea rogue wave measurements up to $h/H_s\sim1.1$, concluding that TF is not suitable to predict larger rogue crest extremes~(see their Fig. 7 in~\cite{donelan2017}). Unfortunately, their analysis is based on a much smaller sampled population of~$\sim10^4$ crest heights and they do not report the confidence bands associated with their probability estimates, nor they provide any parameter values to validate their data analysis.  The deviation of their data from the TF model is most likely due to the small sample of crests. Note also that TF slightly exceeds both the T and F models as an indication that second-order effects are dominant, whereas the linear R model underestimates the return periods. 


For both third- and fourth-order nonlinearities, the return period $N_r$ of a wave whose crest height exceeds the rogue threshold~$1.25H_s\approx11$~m~\cite{DystheKrogstad2008} is nearly~$N_r\sim 10^{4}$ for the El Faro sea state and for the simulated Andrea, Draupner and Killard rogue sea states~\cite{FedeleSP2016}. This is in agreement with oceanic rogue wave measurements~\cite{Christou2014}, which yield roughly the same return period. Similarly, recent measurements off the west coast of Ireland~\cite{Flanagan2016} yield $N_r\sim6\cdot10^4$. In contrast, $N_r\sim 3\cdot 10^{5}$ in a Gaussian sea.

Note that the largest simulated wave crest height exceeds the threshold $1.6H_s\approx14$~m~(see Table~1). This is exceeded on average once every~$10^{6}$ waves in a time series extracted at a point in third- and fourth-order seas and extremely rarely in Gaussian seas, i.e. on average once every~$10^{9}$ waves. This implies that rogue waves observed at a fixed point of the ocean are likely to be rare occurrences of weakly random seas, or Tayfun sea states~\cite{Prestige2015}. Our results clearly confirm that rogue wave generation is the result of the constructive interference (focusing) of elementary waves enhanced by bound nonlinearities in agreement with the theory of stochastic wave groups proposed by Fedele~and~Tayfun~(2009)~\cite{Fedele2009}, which relies on Boccotti's~(2000) theory of quasi-determinism~\cite{Boccotti2000}. Our conclusions are also in agreement with observations~\cite{TayfunFedele2007,Fedele2008,Tayfun2008,Fedele2009}, recent rogue wave analyses~\cite{donelan2017,birkholz2016,FedeleSP2016,dudley2013hokusai} and studies on optical rogue waves caustics analogues~\cite{DudleyCaustics}. 

\subsection*{Time profile of the simulated rogue waves} 

\begin{figure}[h]
\centering\includegraphics[scale=0.4]{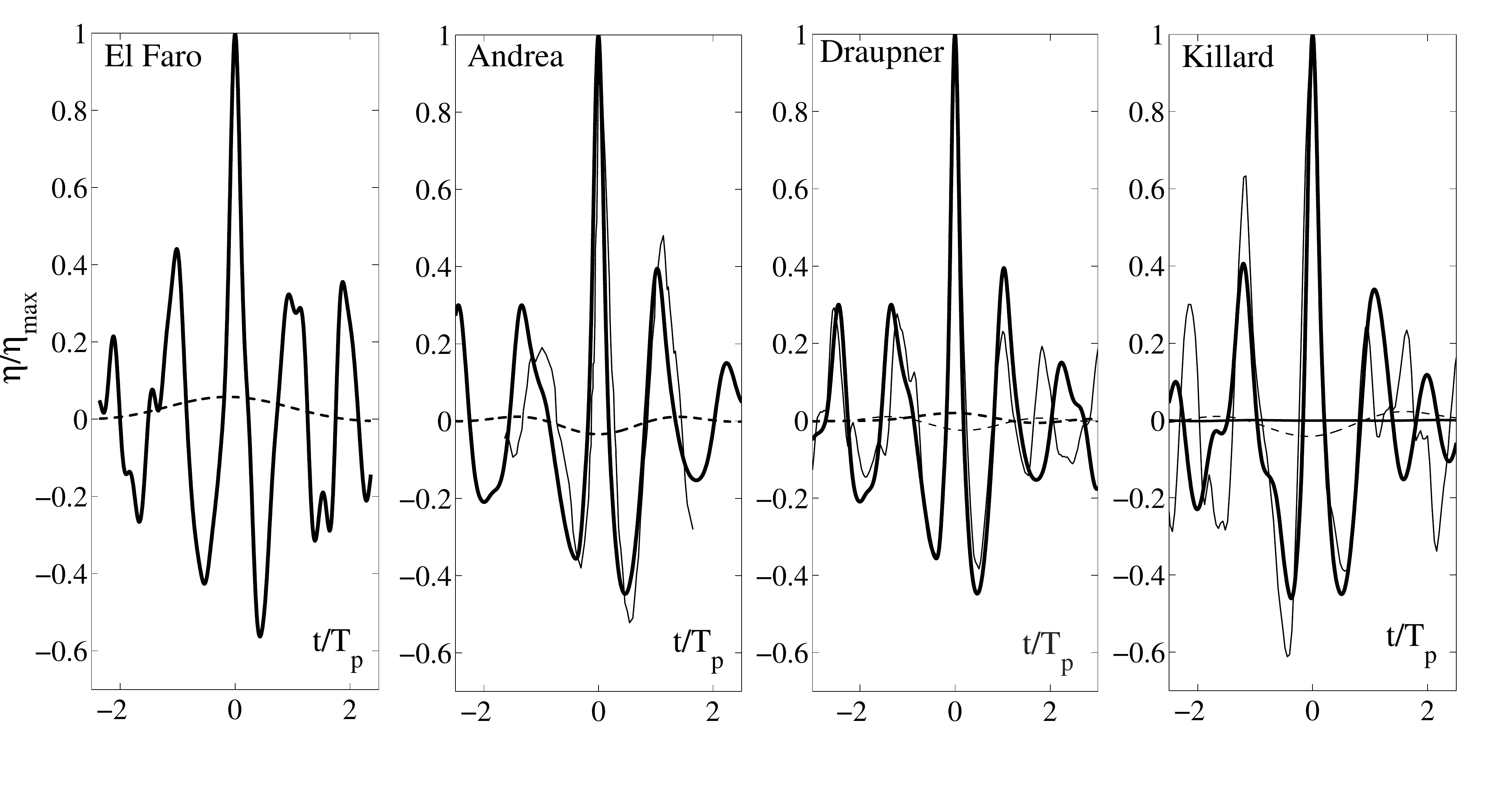} 
\caption{Third-order HOS simulated extreme wave profiles $\eta/\eta_{max}$ (solid) and mean sea levels (MSL)
(dashed) versus the dimensionless time $t/T_p$ for (from left to right) El Faro, Andrea, Draupner and Killard 
waves. $\eta_{max}$ is the maximum crest height given in Table 1. For comparisons, actual measurements (thick 
solid) and MSLs (tick dashed) are also shown for Andrea, Draupner and Killard. Note that the Killard MSL is insignificant and the Andrea MSL is not available. $T_p$ is the dominant wave period~(see Methods section for definitions).
} 
\label{FIG8}
\end{figure}

The wave profile $\eta$ with the largest wave crest height~($>1.6 H_s\approx14$~m) observed in the time series of the surface fluctuations extracted at points randomly sparse over the simulated El Faro domain is shown in the left panel of~Fig.~(\ref{FIG8}). For comparison, the Draupner, Andrea and Killard rogue wave profiles are also shown~\cite{FedeleSP2016}. In the same figure, the mean sea level (MSL) below the crests is also shown. The estimation of the MSL follows by low-pass filtering the measured time series of the wave surface with frequency cutoff $f_c\sim f_p/2$, where $f_p$ is the frequency of the spectral peak~\cite{Walker2004}. 
An analysis of the kinematics~\cite{Fedele_etalJFM2016,FedeleEPL2014} of the simulated rogue waves indicate that such waves were nearly incipient breaking~\cite{Barthelemy2015,BannerSaket2015,Fedele_etalJFM2016} suggesting that larger rogue events are less likely to occur~\cite{Fedele2014,Fedele_etalJFM2016}.  The saturation of the crest height is mainly due to the nonlinear dispersion and it is an energy limiter for rogue waves.

The four wave profiles are very similar suggesting a common generation mechanism of the rogue events. Further, we observe a set-up below the simulated El Faro rogue wave most likely due to the multidirectionality of the sea state. A set-up is also observed for the actual Draupner rogue wave. Indeed, recent studies showed that Draupner occurred in a crossing sea consisting of swell waves propagating at approximately $80$ degrees to the wind sea~\cite{adcock2011did,cavaleri2016draupner}. This would explain the set-up observed under the large wave~\cite{Walker2004} instead of the second-order set-down normally expected~\cite{LonguetHiggins1964}. 


\subsection*{Space-time statistics of the encountered sea state by El Faro before sinking}

The largest crest height of a wave observed in time at a given point of the ocean represents a maximum observed at that point. Clearly, the maximum wave surface height observed over a given area during a time interval, i.e. space-time extreme, is much larger than that observed at a given point. Indeed, in relatively short-crested directional seas such as those generated by hurricanes, it is very unlikely that an observed large crest at a given point in time actually coincides with the largest crest of a group of waves propagating in space-time. In contrast, in accord with Boccotti's (2000) QD theory, it is most likely that the sea surface was in fact much higher somewhere near the measurement point.

\begin{figure}[t]
\centering\includegraphics[scale=0.7]{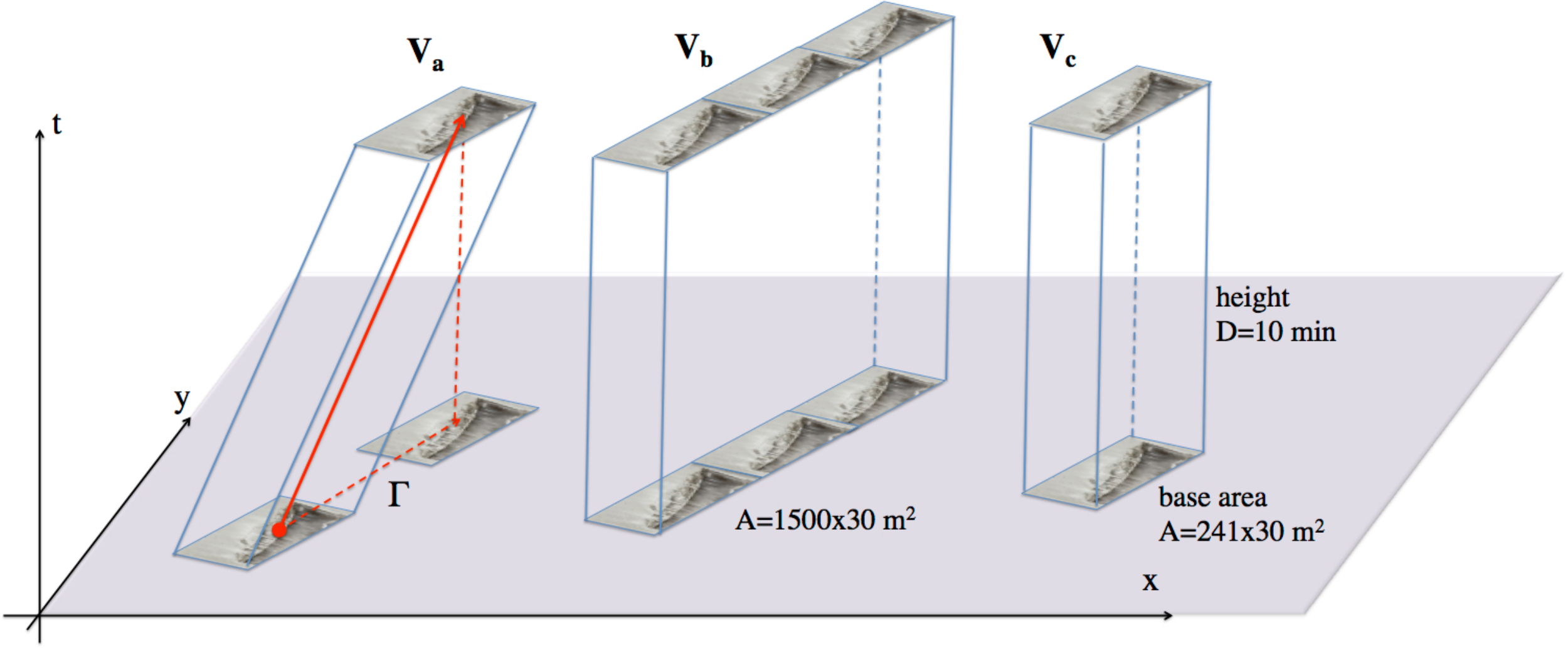} ;
\caption{(Left) the space-time (xyt) volume spanned by the El Faro vessel~(base area $A=241$~x~$30$~$m^2$) while drifting at the speed of $2.5$~$m/s$ over a time interval of $D=10$ minutes along the path $\Gamma$ is that of the slanted parallelepiped $V_a$; (center) the drifting vessel covers the strip area~($1500$~x~$30$~$m^2$) in the 10-minute interval and the associated space-time volume is that of the parallelepiped $V_b$; (right) if the vessel would be anchored at a location for the same duration, it would span instead the spacetime volume of the straight parallelepiped $V_c$. The solid red arrowed line denotes the space-time path of El Faro while drifting along the path $\Gamma$. The vertical axis is time (t) and the other two axes refer to the space dimensions (x) and (y) respectively.}
\label{FIGST}
\end{figure}

Space-time wave extremes can be modeled stochastically~\cite{Fedele2012,fedele2013} drawing on the theory of Euler Characteristics of random fields~\cite{adler1981geometry,adler2009random,adler2000excursion} and nonlinear wave statistics~\cite{TayfunFedele2007}. In the following, we present a new stochastic model for the prediction of space-time extremes~\cite{Fedele2012} that accounts for both second and third-order nonlinearities~\cite{fedele2015JPO}. Drawing on Fedele's work~\cite{Fedele2012,fedele2015JPO} considers a 3-D non-Gaussian field  $\eta(x,y,t)$ in space-time over an area $A$ for a time period of $D$~(see Fig.~\eqref{FIGST}). The area cannot be too large since the wave field may not be homogeneous. The duration should be short so that spectral changes occurring in time are not so significant and the sea state can be assumed as stationary. Then, the third-order nonlinear probability $P_{\mathrm{FST}}^{(nl)}(\xi;A,D)$ that the maximum surface elevation $\eta_{\max}$ over the area $A$ and during the time interval $D$ exceeds the generic threshold $\xi H_{s}$ is equal to the probability of exceeding the threshold $\xi_{0}$, which accounts for kurtosis effects only, that is 
\begin{equation}
\label{Pid2}
P_{\mathrm{FST}}^{(nl)}(\xi;A,D)=P_{\mathrm{ST}}(\xi_{0};A,D) \left(1+\varLambda\xi_{0}^{2}(4\xi_{0}^{2}-1)\right).
\end{equation}

The Gaussian probability of exceedance
\begin{equation}
P_{\mathrm{ST}}(\xi;A,D) =\mathrm{Pr}\left\{ \eta_{\max}>\xi H_{s}\right\} =(16M_{3}\xi^{2}+4M_{2}\xi+M_{1}) P_{\mathrm{R}}(\xi),\label{PV}
\end{equation}
where $P_{\mathrm{R}}(\xi)$ is the Rayleigh exceedance probability of Eq.~\eqref{PR}.

Here, $M_{1}$ and $M_{2}$ are the average number of 1-D and 2-D waves that can occur on the edges and boundaries of the volume $\Omega$, and $M_{3}$ is the average number of 3-D waves that can occur within the volume~\cite{Fedele2012}. These all depend on the directional wave spectrum and its spectral moments $m_{ijk}$ defined in the Methods section.  

The amplitude $\xi$ accounts for both skewness and kurtosis effects and it relates to $\xi_0$ via the Tayfun (1980) quadratic equation 
\begin{equation}
\xi=\xi_{0}+2\mu\xi_{0}^{2}.\label{sub2}
\end{equation}

Given the probability structure of the wave surface defined by Eq.~\eqref{Pid2}, the nonlinear mean maximum surface or crest height $\overline{h}_{\mathrm{FST}}=\xi_{\mathrm{FST}}H_s$ attained over the area $A$ during a time interval $D$ is given, according to Gumbel (1958), by
\begin{equation}
\xi_{\mathrm{FST}}=\overline{h}_{\mathrm{FST}}/H_{s}=\xi_{\mathrm{m}}+2\mu\xi_{\mathrm{m}}^{2}+\frac{\gamma_{e}\left(1+4\mu\xi_{\mathrm{m}}\right)}{16\xi_{\mathrm{m}}-\frac{32M_{3}\xi_{\mathrm{m}}+4M_{2}}{16M_{3}\xi_{\mathrm{m}}^{2}+4M_{2}\xi_{\mathrm{m}}+M_{1}}-\Lambda \frac{2\xi_{\mathrm{m}}(8\xi_{\mathrm{m}}^{2}-1)}{1+\Lambda\xi_{\mathrm{m}}^{2}(4\xi_{m}^{2}-1)}  },\label{xist}
\end{equation}
where the most probable surface elevation value $\xi_{\mathrm{m}}$ satisfies $P_{\mathrm{FST}}(\xi_{\mathrm{m}};A,D)=1$~(see Eq.~\eqref{Pid2}). 
%

The nonlinear mean maximum surface or crest height $h_{\mathrm{T}}$ expected at a point during the time interval $D$ follows from Eq.~\eqref{xist} by setting $M_2=M_3=0$ and $M_1=N_{\mathrm{D}}$, where $N_{\mathrm{D}}=D/\bar{T}$ denotes the number of wave occurring during $D$ and $\bar{T}$ is the mean up-crossing period (see Methods section). The second-order counterpart of the FST model ($\Lambda=0$) has been implemented in WAVEWATCH~III~\cite{barbariol2017}. The linear mean counterpart follows from Eq.~\eqref{xist} by setting $\mu=0$ and $\Lambda=0$.

The statistical interpretations of the probability $P_{\mathrm{FST}}^{(nl)}(\xi;A,D)$ and associated space-time average maximum $\overline{h}_{\mathrm{FST}}$ are as follows. Consider an ensemble of $N$ realizations of a stationary and homogeneous sea state of duration $D$, each of which has similar statistical structure to the El Faro wave field. On this basis, there would be $N$ samples, say $(\eta_{\max}^{(1)},...,\eta_{\max}^{(N)})$ of the maximum surface height $\eta_{\max}$ observed within the area $A$ during the time interval $D$. Then, all the maximum surface heights in the ensemble will exceed the threshold $\overline{h}_{\mathrm{FST}}$. Clearly, the maximum surface height can exceed by far such  average. Indeed, only in a few number of realizations $N\cdot P_{\mathrm{FST}}^{(nl)}(\xi;A,D)$ out of the ensemble of $N$ sea states, the maximum surface height exceeds a threshold $\xi H_s\gg\overline{h}_{\mathrm{FST}}$ much larger than the expected value.  

To characterize such rare occurrences in third-order nonlinear random seas one can consider the threshold $h_q=\xi_q H_s$ exceeded with probability $q$ by the maximum surface height $\eta_{\max}$ over an area $A$ during a sea state of duration $D$.  This satisfies
\begin{equation} 
P_{\mathrm{FST}}^{(nl)}(\xi_q;A,D)=q.\label{PNL1}
\end{equation} 

The statistical interpretation of $h_{q}$ is as follows: the maximum surface height $\eta_{\max}$ observed within the area $A$ during $D$ exceeds the threshold $h_{q}$ only in $q\thinspace N$ realizations of the above mentioned ensemble of $N$ sea states. 

Note that for large areas, i.e. $\ell>>L_{0}$, the $FST$ model as any other similar models available in literature~\cite{piterbarg1995,Juglard2005,forristall2011,forristall2015,cavaleri2016draupner} will overestimate the maximum surface height over an area and time interval because they all rely on Gaussianity. This implies that there are no physical limits on the values that the surface height can attain as the Gaussian model does not account for the saturation induced by the nonlinear dispersion~\cite{Fedele2014} of ocean waves or wave breaking. Thus, the larger the area $A$ or the time interval $D$, the greater the number of waves sampled in space-time, and unrealistically large amplitudes are likely to be sampled in a Gaussian or weakly nonlinear Gaussian sea. 

\begin{figure}[t]
\centering\includegraphics[scale=0.45]{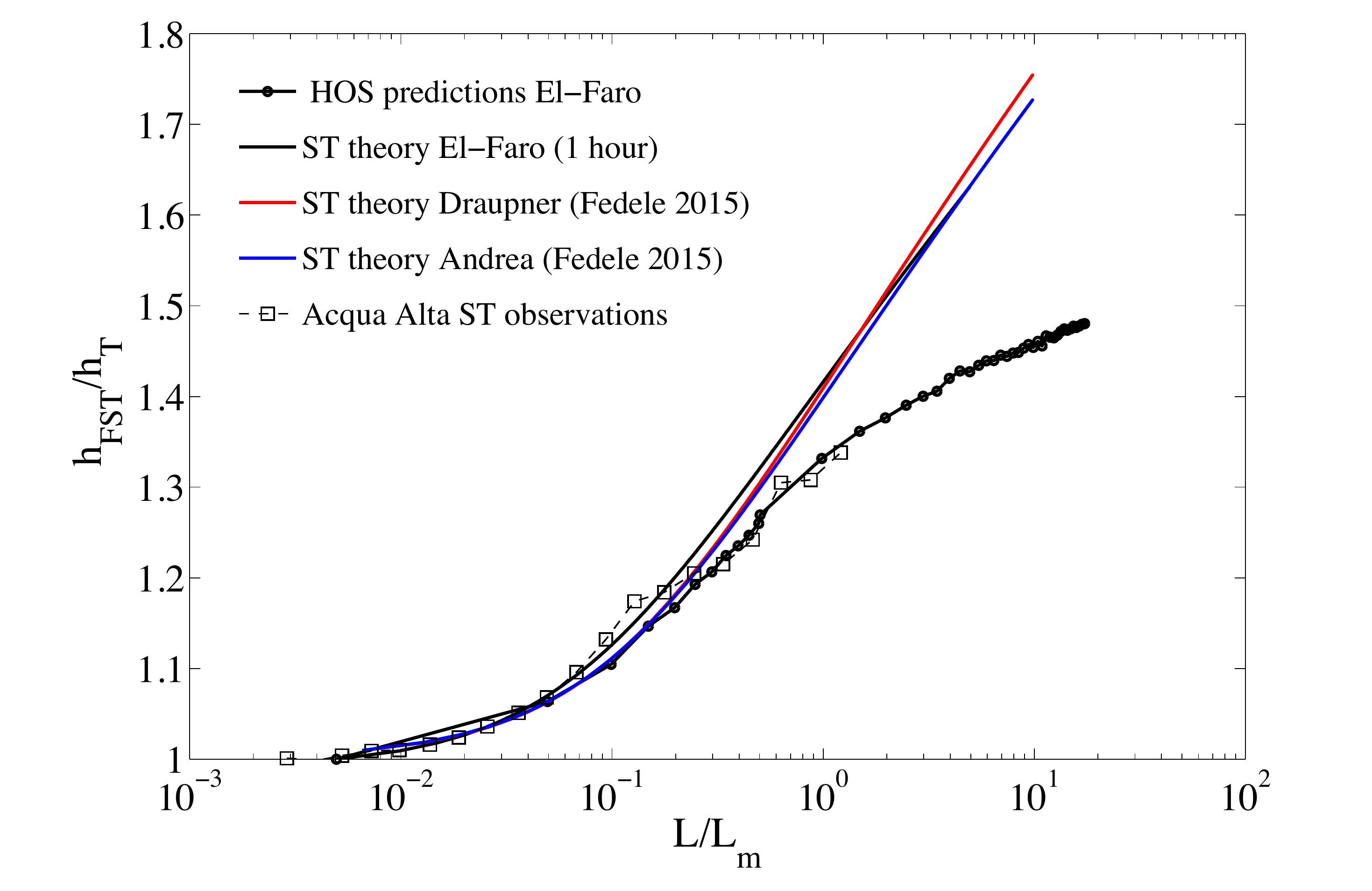} \protect\caption{Space-time extremes: theoretical FST ratios $\overline{h}_{\mathrm{FST}}/\overline{h}_{\mathrm{T}}$ as a function of the area width $\ell/L_0$ for El Faro (black), Draupner (red) and Andrea (blue) sea states,  where $\overline{h}_{\mathrm{FST}}$ is the mean maximum surface height expected over the area $\ell^2$ during a sea state of duration $D=1$ hours and $\overline{h}_{\mathrm{T}}$ is the mean maximum surface height expected at a point. For comparisons, the empirical FST ratio from the El Faro HOS simulations (dashed line) together with the experimental observations at the Acqua Alta tower (squares) are also shown~\cite{fedele2013}. $L_0$ is the mean wavelength.}\label{FIG10}
\end{figure}

This point is elaborated further and demonstrated explicitly by way of the results displayed in Fig.~(\ref{FIG10}). Here, the theoretical (FST) ratio $\overline{h}_{\mathrm{FST}}/\overline{h}_{\mathrm{T}}$ as a function of the area width $\ell/L_0$ is shown for the El Faro, Draupner and Andrea sea states respectively. The FST ratios for Draupner and Andrea are estimated using the European Reanalysis (ERA)-interim data~\cite{fedele2015JPO}. For comparisons, the empirical FST ratio from the El Faro HOS simulations together with the experimental observations at the Acqua Alta tower~\cite{fedele2013} are also shown.  Recall that $\overline{h}_{\mathrm{FST}}$  is the mean maximum surface height expected over the area $\ell^2$ during a sea state of duration $D=1$ hour and $\overline{h}_{\mathrm{T}}$ is the mean maximum surface height expected at a point. Clearly, the theoretical FST ratio for El Faro fairly agrees with the HOS simulations for small areas ($\ell\le L_0$), whereas it yields overestimation over larger areas. We argue that the saturation of the HOS FST ratio over larger areas is an effect of the nonlinear dispersion which is effective in limiting the wave growth as a precursor to breaking~\cite{Fedele2014,Fedele_etalJFM2016}.

Note that the FST ratios for all the three sea states are nearly the same for $\ell\leq L_0$. These results are very encouraging as they suggest possible statistical similarities and universal laws for space-time extremes in wind sea states. Moreover, for $\ell\sim L_{0}$ the mean wave surface maximum expected over the area is 1.35 times larger than that expected at a point in agreement with Acqua Alta sea observations~\cite{fedele2013}.


\subsection*{The occurrence frequency of a rogue wave by the El Faro vessel} 

The data suggests that the El Faro vessel was drifting at an average speed of approximately~$2.5$~m/s prior to its sinking. This is considered in our analysis as follows. First, define the two events $R=\text{"El Faro encounters a rogue wave along its navigation route"}$ and  $S=\text{"El Faro sinks"}$. We know that the event $S$ happened. As a result, one should consider the conditional probability
\begin{equation}
\mathrm{Pr}[R|S]=\frac{\mathrm{Pr}[S|R]\cdot\mathrm{Pr}[R]}{\mathrm{Pr}[S]}.
\label{PRS}
\end{equation}
Here, $\mathrm{Pr}[S]$ is the unconditional probability of the event that El Faro sinks. This could be estimated from worldwide statistics of sunk vessels with characteristics similar to El Faro. $\mathrm{Pr}[S|R]$ is the conditional probability that El Faro sinks given that the vessel encountered a rogue wave. This probability can be estimated by Monte Carlo simulations of the nonlinear interaction of the vessel with the rogue wave field. 

Our rogue wave analysis provides an estimate of the unconditional probability~$\mathrm{Pr}[R]$ that El Faro encounters a rogue wave along its navigation or drifting route by means of the exceedance probability, or occurrence frequency~$P_e(h)$. This is the probability that a vessel along its navigation path encounters a rogue wave whose crest height exceeds a given threshold $h$. 
The encounter of a rogue wave by a moving vessel is analogous to that of a big wave that a surfer is in search of. His likelihood to encounter a big wave increases if he moves around a large area instead of staying still. 
This is a space-time effect which is very important for ship navigation and must be accounted for~\cite{Pierson1953,Lindgren1999,Rychlik2000,Fedele2012}. 

The exceedance probability $P_e(h)$ is formulated as follows. Consider a random wave field whose surface elevation at a given point $(x,y)$ in a fixed frame at time $t$ is $\eta(x,y,t)$.  Consider a vessel of area $A$ that navigates through the wave field at a constant speed $V$ along a straight path at an angle $\beta$ with respect to the $x$ axis. Define also $(x_e,y_e)$ as a cartesian frame moving with the ship. Then, the line trajectories of any point $(x_e,y_e)$ of the vessel in the fixed frame are given by 
\begin{equation} 
x=x_e+V\cos(\beta)t,\quad y=y_e+V\sin(\beta)t,\label{xy}
\end{equation} 
where for simplicity we assume that at time $t=0$ the center of gravity of the vessel is at the origin of the fixed frame. 

The surface height $\eta_c(t)$ encountered by the moving vessel, or equivalently the surface fluctuations measured by a wave probe installed on the ship, is
\begin{equation} 
\eta_c(x_e,y_e,t)=\eta(x_e+V\cos(\beta)t,y_e+V\cos(\beta)t,t),\label{etac}
\end{equation} 
If $\eta$ is a Gaussian wave field homogeneous in space and stationary in time, then so is $\eta_c$ with respect to the moving frame $(x_e,y_e,t)$. The associated space-time covariance is given by
\begin{equation} 
\Psi(X,Y,T)=\overline{\eta_c(x_e,y_e,t) \eta_c(x_e+X,y_e+Y,t+T)}=\int S(f,\theta)\cos(k_{x}X+k_{y}Y-2\pi f_{e}T) \mathrm{d}f\mathrm{d}\theta,\label{etac_psi}
\end{equation}
where $k_x=k\cos(\theta)$, $k_y=k\sin(\theta)$ and $k$ is the wavenumber associated with the frequency $f$ by way of the wave dispersion relation. As a result of the Doppler effect, the encountered, or apparent frequency is~\cite{Pierson1953,Lindgren1999,Rychlik2000}
\begin{equation} 
f_{e}=f-k V\cos(\theta-\beta)/(2\pi),
\end{equation}
and $S(f,\theta)$ is the directional wave spectrum of the sea state. Note that when the vessel moves faster than waves coming from a direction $\theta$, the apparent frequency $f_{e}<0$ and for an observer on the ship waves appear to move away from him/her. In this case, the direction of those waves should be reversed~\cite{Pierson1953}, i.e. $\theta=\theta+\pi$, and $f_e$ set as positive. 

The spectral moments $m_{ijk}^{(e)}$ of the encountered random field readily follow from the coefficients of the Taylor series expansion of $\Psi(X,Y,T)$ around $(X=0,Y=0,T=0)$. In particular,
\begin{equation} 
m_{ijk}^{(e)}=\frac{\partial^{i+j+k}\Psi}{\partial X^i \partial Y^j \partial T^k}\Big|_{X=Y=T=0}=\int S(f,\theta)k_{x}^{i}k_{y}^{j}f_{e}^{k}\mathrm{d}f\mathrm{d}\theta.
\end{equation} 

The nonlinear space-time statistics can then easily processed by using the encountered spectral moments $m_{ijk}^{(e)}$ using the FST model~\cite{Fedele2012,fedele2015JPO}, which is based on~Eq.~\eqref{Pid2} as described above. Note that for generic navigation routes the encountered wave field $\eta_c$ is a non-stationary random process of time. Thus, the associated spectral moments will vary in time. The space-time statistics can be still computed by first approximating the route by a polygonal made of piecewise straight segments along which the random process $\eta_c$ is assumed as stationary. 

\begin{figure}[t]
\centering\includegraphics[scale=0.75]{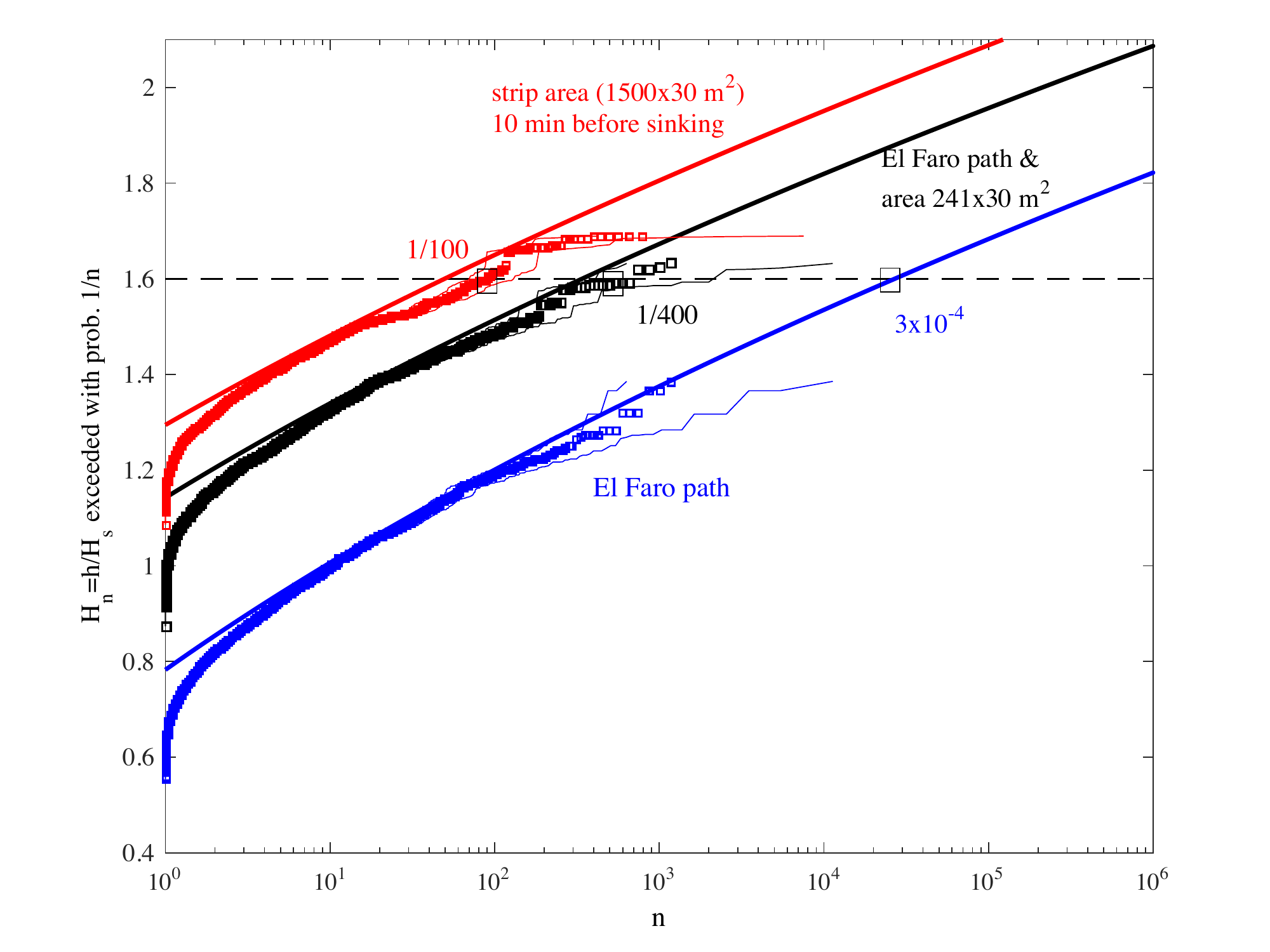} 
\caption{HOS (squares) and theoretical (solid lines) predictions for the normalized nonlinear threshold $h_n/H_s$ exceeded with probability $1/n$; i) along the straight path $\Gamma$ spanned by El Faro while drifting at an estimated approximate average speed of~$2.5$~m/s over a time interval of 10 minutes (blue), ii) and also accounting for the vessel size~($241$~x~$30$~$m^2$)  (black), and over the strip area~($1500$~x~$30$~$m^2$) spanned by the vessel in a 10-minute interval (red). Confidence bands are also shown~(light dashes). Horizontal line denotes the threshold~$1.6H_s\approx14$~m, which is exceeded with probability~$3\cdot10^{-4}$,$1/400$ and~$1/100$ for the three cases shown.}
\label{FIG11}
\end{figure}

Fig.~\eqref{FIG11} illustrates the HOS and theoretical predictions for the normalized nonlinear threshold $h_n/H_s$ exceeded with probability $1/n$, where $n$ is the number of waves. In particular, consider an observer on the vessel moving along the straight path $\Gamma$ spanned by El Faro drifting against the dominant sea direction over a time interval of 10 minutes. In space-time the observer spans the solid red line shown in Fig.~\eqref{FIGST}. In this case, he has a probability~$P_e\sim 3\cdot10^{-4}$ to encounter a wave whose crest height exceeds the threshold~$1.6H_s\approx14$~m~(blue lines). If we also account for the vessel size~(base area $A=241$~x~$30$~$m^2$), in space-time El Faro spans the volume of the slanted parallelepiped $V_a$ shown in Fig.~\eqref{FIGST}. In this case, the exceedance probability $P_e(V_a)$ further increases to~$1/400$~(black lines). Note that If the vessel would be anchored at a location for the same duration, in spacetime it would span instead the volume of the vertical parallelepiped $V_c$ shown in the same Figure. Note that the two parallelepipeds cover the same space-time volume $A$~x~$D$, with the base area $A$ and height $D=10$~$min$. For the case of the anchored vessel, the associated exceedance probability $P_e(V_c)$ is roughly the same as $P_e(V_a)$ since El Faro was drifting at a slow speed. Larger drift speeds yield larger $P_e(V_a)$ since the vessel encounters waves more frequently than if it was anchored, because of the Doppler effect~\cite{Lindgren1999,Rychlik2000}. Moreover, the drifting vessel covers the strip area~($1500$~x~$30$~$m^2$) in the 10-minute interval and the associated space-time volume is that of the parallelepiped $V_b$ shown in Fig.~\eqref{FIGST}, which has a larger volume than that of $V_a$. As a result, the occurrence frequency $P_e(V_b)$ of a rogue wave associated with $V_b$ is larger and it increases to~$\sim 1/100$~(see red lines in Fig.~\eqref{FIG11}). However, El Faro does not visit the entire volume $V_b$, but it only spans the smaller volume $V_a$. Thus, the conditional probability $P_e(V_a|V_b)$ that the drifting El Faro encounters a rogue wave given that a rogue wave occurred over the larger spacetime volume $V_b$ is $P_e(V_a)/P_e(V_b)\sim1/4$. Furthermore, a fixed observer has a much lower probability~$P_e\sim10^{-6}$ to pick randomly from a time series extracted at a point a wave whose crest height exceeds ~$1.6H_s$~(see Fig.~\ref{FIG7}, TF model, red solid line).
Finally, we observed that the exceedance probability~$P_e(V_a)$ for the drifting El Faro does not scale linearly with time because of nonlinearities that reduce the natural dispersion of waves. Indeed, assuming that El Faro drifts over a time interval 5 times longer ($50$ minutes), $P_e(V_a)$ just increases roughly by 3 times,~$\sim1/130$. 

\begin{figure}[h]
\centering\includegraphics[scale=0.4]{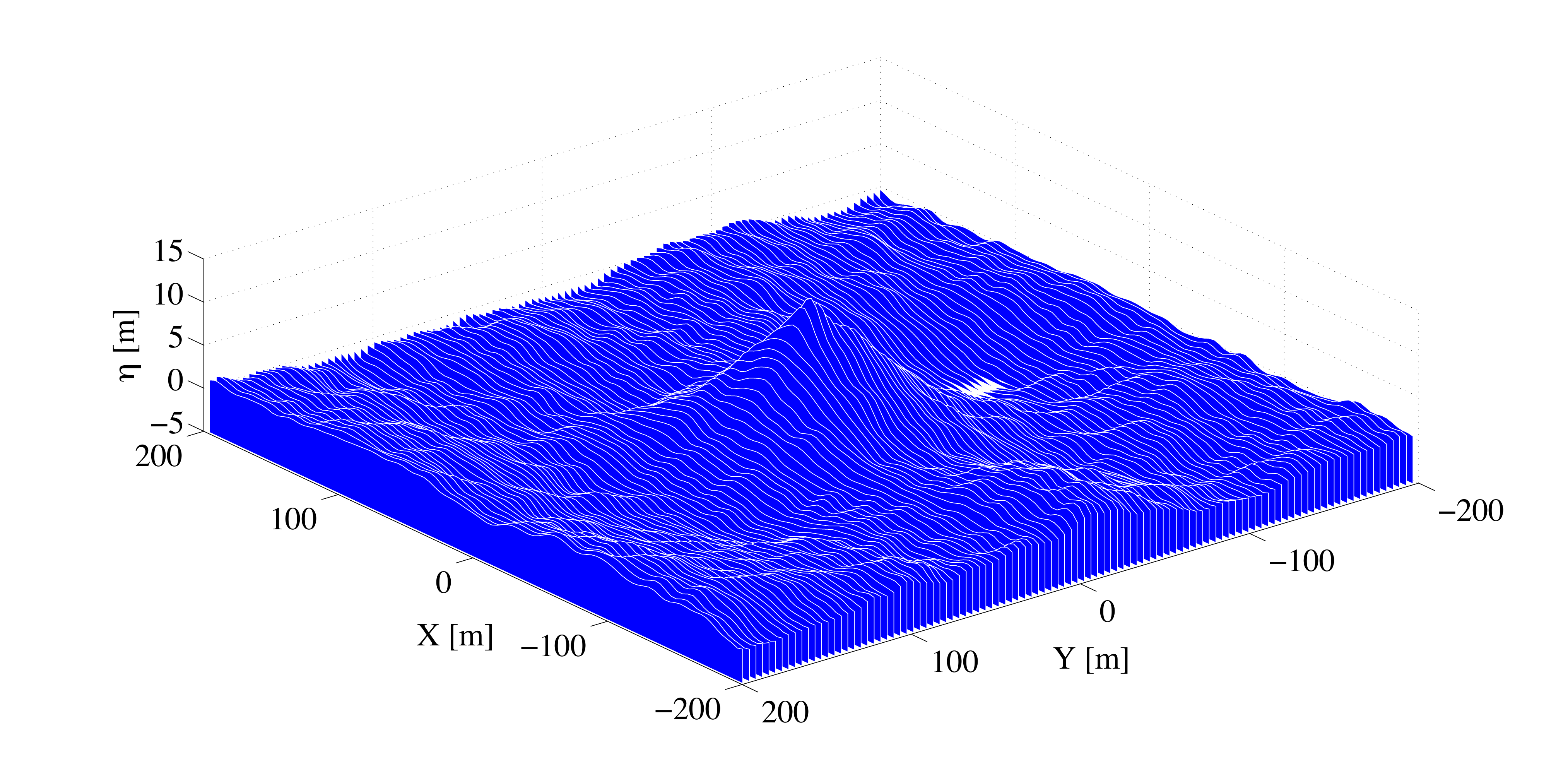} \protect\caption{HOS simulations: expected spatial shape of a rogue wave whose crest height is~$>1.6 H_s\approx14$~m.}\label{FIG12}
\end{figure}

\section*{Discussions}

Our present studies open a new research direction on the prediction of rogue waves during hurricanes. Indeed, the impact of our studies is two-fold. On the one hand, the present statistical analysis provides the basis for an improved understanding of how rogue waves originate during hurricanes. On the other hand, the proposed stochastic model for the encounter probability of a rogue wave provides the basis in the next generation of wave forecast models for a predictive capability of wave extremes and early warnings for shipping companies and others to avoid dangerous areas at risk of rogue waves.

\section*{Methods}

\subsection*{Wave parameters}\label{Ss_Wp}

The significant wave height $H_s$ is defined as the mean value $H_{1/3}$ of the highest one-third of wave heights. It can be estimated either from a zero-crossing analysis or more easily from the wave omnidirectional spectrum $S_o(f)=\int_{0}^{2\pi} S(f,\theta)\mathrm{d}{\theta}$  as $H_s \approx 4\sigma$, where $\sigma=\sqrt{m_0}$ is the standard deviation of surface elevations, $m_j=\int S_o(f) f^j\mathrm{d}f$ are spectral moments. Further, $S(f,\theta)$ is the directional wave spectrum with $\theta$ as the direction of waves at frequency $f$, and the cyclic frequency is $\omega=2\pi f$. 

The dominant wave period $T_{p}=2\pi/\omega_p$ refers to the cyclic frequency $\omega_p$ of the spectral peak. The mean zero-crossing wave period $T_{0}$ is equal to $2\pi/\omega_0$, with $\omega_0=\sqrt{m_2/m_0}$. The associated wavelength $L_{0}=2\pi/k_0$ follows from the linear dispersion relation $\omega_0 = \sqrt{gk_0 \tanh(k_0 d)}$, with $d$ the water depth. The mean spectral frequency is defined as~$\omega_{m}=m_{1}/m_{0}$~\cite{Tayfun1980} and the associated mean period $T_m$ is equal to $2\pi/\omega_m$.  A characteristic wave steepness is defined as $\mu_m=k_m\sigma$, where $k_{m}$ is the wavenumber corresponding to the mean spectral frequency $\omega_{m}$~\cite{Tayfun1980}. The following quantitites are also introduced: $q_m = k_m d, Q_m = \tanh q_m$, the phase velocity $c_m = \omega_m/k_m$, the group velocity $c_g=c_m\left[1+2q_{m}/\mathrm{sinh(2}q_{m})\right]/2$. 

The spectral bandwidth $\nu=(m_0 m_2/m_1^2-1)^{1/2}$ gives a measure of the frequency spreading. The angular spreading~$\sigma_{\theta}=\sqrt{\int_0^{2\pi}D(\theta)(\theta-\theta_m)^2 \mathrm{d}\theta}$, where $D(\theta)=\int_0^{\infty}S(\omega,\theta)\mathrm{d}\omega/\sigma^2$ and $\theta_m=\int_0^{2\pi}D(\theta)\theta\mathrm{d}\theta$ is the mean direction. Note that~$\omega_0=\omega_m\sqrt{1+\nu^2}$.

The wave skewness $\lambda_3$ and the excess kurtosis $\lambda_{40}$ of the zero-mean surface elevation $\eta(t)$ are given by
\begin{equation}
\lambda_3=\overline{\eta^3}/\sigma^3,\qquad\lambda_{40}=\overline{\eta^4}/\sigma^4-3 \,.
\end{equation}
Here, overbars imply statistical averages and $\sigma$ is the standard deviation of surface wave elevations.  

For second-order waves in deep water~\cite{Fedele2009}
\begin{equation}
\lambda_{3}\approx 3\mu_m(1-\nu+\nu^2),
\end{equation}
and the following bounds hold~\cite{Tayfun2006}
\begin{equation}
3\mu_m(1-\sqrt{2}\nu+\nu^2) \leq \lambda_3 \leq 3\mu_m.
\end{equation}
Here, $\nu$ is the spectral bandwidth defined above and the characteristic wave steepness $\mu_m=k_m\sigma$, where $k_{m}$ is the wavenumber corresponding to the mean spectral frequency $\omega_{m}$~\cite{Tayfun1980}. For narrowband (NB) waves, $\nu$ tends to zero and the associated skewness $\lambda_{3,NB}=3\mu_m$~\cite{Tayfun1980,TayfunFedele2007,Fedele2009}.


For third-order nonlinear random seas the excess kurtosis
\begin{equation}
\lambda_{40}=\lambda_{40}^{d}+\lambda_{40}^{b}\label{C4total}
\end{equation} 
comprises a dynamic component $\lambda_{40}^{d}$ due to nonlinear quasi-resonant
wave-wave interactions~\cite{Janssen2003,Janssen2009} and a Stokes bound harmonic contribution
$\lambda_{40}^{b}$~\cite{JanssenJFM2009}. 
In deep water it reduces to the simple form~$\lambda_{40,NB}^{b}=18\mu_m^{2}=2\lambda_{3,NB}^2$~\cite{Janssen2009,JanssenJFM2009,Janssen2014a} where $\lambda_{3,NB}$ is the skewness of narrowband waves~\cite{Tayfun1980}.

As for the dynamic component, Fedele~\cite{fedele2015kurtosis} recently revisited Janssen's~\cite{Janssen2003} weakly nonlinear formulation for $\lambda_{40}^{d}$. In deep water, this is given in terms of a six-fold integral that depends on the Benjamin-Feir index $BFI=\mu_m/\sqrt{2}\nu$ and the parameter $R=\sigma_{\theta}^{2}/2\nu^{2}$, which is a dimensionless measure of the multidirectionality of dominant waves~\cite{Janssen2009,Mori2011}. 
As waves become unidirectional (1D) waves $R$ tends to zero.
 
\subsection*{The Tayfun-Fedele model for crest heights} 
 
We define $P(\xi)$ as the probability that a wave crest observed at a fixed point of the ocean in time exceeds the threshold $\xi H_s$. For weakly nonlinear nonlinear seas, this probability can be described by the third-order Tayfun-Fedele model~\cite{TayfunFedele2007}, 
\begin{equation}
P_{TF}(\xi)=\mathrm{Pr}\left[h>\xi\,H_s\right]=\mathrm{exp}\left(-8\,\xi_{0}^{2}\right)\left[1+\varLambda \xi_{0}^{2}\left(4\,\xi_{0}^{2}-1\right)\right],\label{Pid}
\end{equation}
where $\xi_{0}$ follows from the quadratic equation $\xi=\xi_{0}+2\mu\,\xi_{0}^{2}$~\cite{Tayfun1980}.
Here, the Tayfun wave steepness~$\mu=\lambda_{3}/3$ is of $O(\mu_m)$ and it is a measure of second-order bound nonlinearities as it relates to the skewness $\lambda_{3}$ of surface elevations~\cite{Fedele2009}.  The parameter~$\varLambda=\lambda_{40}+2\lambda_{22}+\lambda_{04}$ is a measure of third-order nonlinearities and is a function of the fourth order cumulants $\lambda_{nm}$ of the wave surface $\eta$ and its Hilbert transform $\hat{\eta}$~\cite{TayfunFedele2007}. In particular, $\lambda_{22}=\overline{\eta^2\hat{\eta}^2}/\sigma^4-1$~and~$\lambda_{04}=\overline{\hat{\eta}^4}/\sigma^4-3$. 
In our studies $\varLambda$ is approximated solely in terms of the excess kurtosis as~$\varLambda_{\mathrm{appr}}={8\lambda_{40}}/{3}$ by assuming the relations between cumulants~\cite{Janssen2006} $\lambda_{22}=\lambda_{40}/3$ and $\lambda_{04}=\lambda_{40}$. These, to date, have been proven to hold for linear and second-order narrowband waves only \cite{TayfunLo1990}. For third-order nonlinear seas, our numerical studies indicate that $\varLambda\approx\varLambda_{\mathrm{appr}}$ within a $3\%$ relative error in agreement with observations~\cite{fedeleNLS,TayfunOMAE2007}.

For second-order seas, referred to as Tayfun sea states~\cite{Prestige2015}, $\varLambda=0$ only and $P_{TF}$ in Eq.~\eqref{Pid} yields the Tayfun~(T)~distribution~\cite{Tayfun1980}
\begin{equation}
P_{T}(\xi)=\mathrm{exp}\left(-8{\xi_{0}^2}\right).\label{PT}
\end{equation}
For Gaussian seas, $\mu =0$ and $\varLambda=0$ and $P_{TF}$ reduces to the Rayleigh~(R)~distribution
\begin{equation}
P_{R}(\xi)=\mathrm{exp}\left(-8{\xi^{2}}\right).\label{PR}
\end{equation}

Note that the Tayfun distribution represents an exact result for large second order wave crest heights and it depends solely on the steepness parameter defined as $\mu=\lambda_{3}/3$~\cite{Fedele2009}. 

\subsection*{The Forristall model}
The exceedance probability is given by~\cite{Forristall2000} 
\begin{equation}
P_{F}(\xi)=\mathrm{exp}\left(-{(\xi/\alpha)^{\beta}}\right),\label{F}
\end{equation}
where $\alpha=0.3536+0.2561 S_1+0.0800 U_r$, $\beta=2-1.7912 S_1-0.5302 U_r+0.284 U_r^2$ for multi-directional (short-crested) seas. Here, $S_1=2\pi H_s/(g T_m^2)$ is a characteristic wave steepness and the Ursell number $U_r=H_s/(k_m^2 d^3)$, where $k_m$ is the wavenumber associated with the mean period $T_m=m_0/m_1$ and $d$ is the water depth.

\subsection*{Space-Time Statistical Parameters}\label{Ss_STSP}

For space-time extremes, the coefficients in Eq.~\eqref{PV} are given by~\cite{Baxevani2006,Fedele2012}
\[
M_{3}=2\pi\frac{D}{\overline{T}}\frac{\ell_{x}}{\overline{L_{x}}}\frac{\ell_{y}}{\overline{L_{y}}}\alpha_{xyt},
\]
\[
M_{2}=\sqrt{2\pi}\left(\frac{D}{\overline{T}}\frac{\ell_{x}}{\overline{L_{x}}}\sqrt{1-\alpha_{xt}^{2}}+\frac{D}{\overline{T}}\frac{\ell_{y}}{\overline{L_{y}}}\sqrt{1-\alpha_{yt}^{2}}+\frac{\ell_{x}}{\overline{L_{x}}}\frac{\ell_{y}}{\overline{L_{y}}}\sqrt{1-\alpha_{xy}^{2}}\right),
\]
\[
M_{1}=N_{D}+N_{x}+N_{y},
\]
where 
\[
N_{D}=\frac{D}{\overline{T}},\qquad N_{x}=\frac{\ell_{x}}{\overline{L_{x}}},\qquad N_{y}=\frac{\ell_{y}}{\overline{L_{y}}}
\]
are the average number of waves occurring during the time interval
D and along the x and y sides of length $\ell_{x}$ and $\ell_{y}$
respectively. They all depend on the mean period $\overline{T}$,
mean wavelengths $\overline{L_{x}}$ and $\overline{L_{y}}$ in $x$
and $y$ directions: 
\[
\overline{T}=2\pi\sqrt{\frac{m_{000}}{m_{002}}},\qquad\overline{L_{x}}=2\pi\sqrt{\frac{m_{000}}{m_{200}}},\qquad\overline{L_{y}}=2\pi\sqrt{\frac{m_{000}}{m_{020}}}
\]
and
\[
\alpha_{xyt}=\sqrt{1-\alpha_{xt}^{2}-\alpha_{yt}^{2}-\alpha_{xy}^{2}+2\alpha_{xt}\alpha_{yt}\alpha_{xy}}.
\]
Here, 
\[
m_{ijk}=\iint k_{x}^{i}k_{y}^{j}f^{k}S(f,\theta)dfd\theta
\]
are the moments of the directional spectrum $S(f,\theta)$ and 
\[
\alpha_{xt}=\frac{m_{101}}{\sqrt{m_{200}m_{002}}},\qquad\alpha_{yt}=\frac{m_{011}}{\sqrt{m_{020}m_{002}}},\qquad\alpha_{xy}=\frac{m_{110}}{\sqrt{m_{200}m_{020}}}.
\]

\subsection*{The Higher Order Spectral (HOS) method}\label{Ss_HOS}
 
The HOS, developed independently by Dommermuth \& Yue~\cite{DommermuthYue1987HOS}  and West 
\textit{et al.} \cite{West1987} is a numerical pseudo-spectral method, based on a perturbation expansion of the wave potential function up to a prescribed order of nonlinearities $M$ in terms of a small parameter, the characteristic wave steepness. The method solves for nonlinear wave-wave interactions up to the specified order $M$ of a number $N$ of free waves (Fourier modes). The associated boundary value problem is solved by way of a pseudo-spectral technique, ensuring a computational cost which scales linearly with $M^2N \log(N)$~\cite{Fucile_PhD,Schaffer}. As a result, high computational efficiency is guaranteed for simulations over large spatial domains. In our study we used the West formulation~\cite{West1987}, which accounts for all the nonlinear terms at a given order of the perturbation expansion. The details of the specific algorithm are given in Fucile~\cite{Fucile_PhD} and Fedele \textit{et al.}~\cite{fedele2016prediction}. The wave field is resolved using $1024$~x~$1024$ Fourier modes on a spatial area of $4000$m~x~$4000$m. Initial conditions for the wave potential and surface elevation are specified from the directional spectrum as an output of WAVEWATCH III~\cite{tolman2014}. 

\section*{Data Availability}
All the publicly available data and information about the El Faro accident are posted on the National Transportation Safety Board (NTSB) website~\cite{elfaro}.

\hspace{2cm}
\bibliographystyle{unsrt}
\bibliography{biblioFranco}

\section*{Acknowledgments}
This manuscript is based on a study on the prediction of rogue waves during Hurricane Joaquin provided as a supplement to the National Transportation Safety Board (NTSB) to assist them in their investigation of the sinking of the Merchant Vessel El Faro, which occurred east of the Bahamas on October 1, 2015. The authors thank Emilio F. Campana for his support and incisive intellectual discussions and Fabio Fucile for helping with the HOS simulations. 

C. Lugni was supported by the Research Council of Norway through the Centres of Excellence funding scheme AMOS, project number 223254 and by the Flagship Project RITMARE - The Italian Research for the Sea - coordinated by the Italian National Research Council.

\section*{Author contributions statement}

The concept and design was provided by F. Fedele, who coordinated the scientific effort together with C. Lugni. C. Lugni performed numerical simulations and developed specific codes for the analysis. The wave statistical analysis was performed by F. Fedele together with C. Lugni. The overall supervision was provided by F. Fedele; A. Chawla performed the WAVEWATCH simulations and made ongoing incisive intellectual contributions. All authors participated in the analysis and interpretation of results and the writing of the manuscript.

\section*{Additional information}

Competing financial interests: The authors declare no competing financial interests.

\section*{Figure Legends}

\begin{description}

\item[Figure 1] WAVEWATCH III parameters history during Hurricane Joaquin around the location where the El Faro vessel sank. (top-left) Hourly variation of the significant wave height $H_s$, (top-right) dominant wave period $T_p$, (bottom-left) dominant wave direction and (bottom-right) normalized $U_{10}/U_{10,max}$ wind speed (solid line) and direction (dashed line). Maximum wind speed $U_{10,max}=51 m/s$. Red vertical lines delimit the 1--hour interval during which the El Faro vessel sank.

\item[Figure 2] WAVEWATCH III parameters history during Hurricane Joaquin around the location where the El Faro vessel sank. (top) Hourly variation of the spectral bandwidth $\nu$ history, (center) directional spreading $\theta_v$ and (bottom) directional factor $R=\frac{1}{2}\nu^2/\theta_v^2$. Red vertical lines delimit the 1-hour interval during which the El Faro vessel sank.

\item[Figure 3] WAVEWATCH III parameters history during Hurricane Joaquin around the location where the El Faro vessel sank. (top) Hourly variation of the Tayfun steepness $\mu$ (solid line) with bounds (dashed lines), (center) excess kurtosis $\lambda_{40}$ and (bottom) nonlinear coefficient $\Lambda\sim 8\lambda_{40}/3$. Red vertical lines delimit the 1-hour interval during which the El Faro vessel sank.

\item[Figure 4] WAVEWATCH III hindcast directional spectrum ${S}(f,\theta)$~$[m^2 s/rad]$ at approximately the time and location of the El-Faro sinking.

\item[Figure 5] HOS simulations of the El Faro sea state: predicted wavenumber-frequency spectrum~$S(k,\omega)$~$[m^2 s/rad]$. Sea state duration of 1 hour over an area of~$4$~km~x~$4$~km; the wave field is resolved using $1024$~x~$1024$ Fourier modes.

\item[Figure 6] HOS simulations of the El Faro sea state. Crest height scaled by the significant wave height ($\xi$) versus conditional return period ($N_h$) for the (left) Andrea, (center) Draupner and (right) Killard rogue sea states:  HOS numerical predictions ($\square$) in comparison with theoretical models: F=Forristall (blue dashed), T=second-order Tayfun (blue solid), TF=third-order (red solid) and R=Rayleigh distributions (red solid). Confidence bands are also shown~(light dashes). $N_h(\xi)$ is the inverse of the exceedance probability $P(\xi)=\mathrm{Pr}[h>\xi H_s]$. Horizontal lines denote the rogue threshold~$1.25H_s$~\cite{DystheKrogstad2008} and~$1.6H_s$.

\item[Figure 7] Third-order HOS simulated extreme wave profiles $\eta/\eta_{max}$ (solid) and mean sea levels (MSL) (dashed) versus the dimensionless time $t/T_p$ for (from left to right) El Faro, Andrea, Draupner and Killard waves. $\eta_{max}$ is the maximum crest height given in Table 1. For comparisons, actual measurements (thick solid) and MSLs (tick dashed) are also shown for Andrea, Draupner and Killard. Note that the Killard MSL is insignificant and the Andrea MSL is not available. $T_p$ is the dominant wave period~(see Methods section for definitions).

\item[Figure 8] (Left) the space-time (xyt) volume spanned by the El Faro vessel~(base area $A=241$~x~$30$~$m^2$) while drifting at the speed of $2.5$~$m/s$ over a time interval of $D=10$ minutes along the path $\Gamma$ is that of the slanted parallelepiped $V_a$; (center) the drifting vessel covers the strip area~($1500$~x~$30$~$m^2$) in the 10-minute interval and the associated space-time volume is that of the parallelepiped $V_b$; (right) if the vessel would be anchored at a location for the same duration, it would span instead the spacetime volume of the straight parallelepiped $V_c$. The solid red arrowed line denotes the space-time path of El Faro while drifting along the path $\Gamma$. The vertical axis is time (t) and the other two axes refer to the space dimensions (x) and (y) respectively.

\item[Figure 9] Space-time extremes: theoretical FST ratios $\overline{h}_{\mathrm{FST}}/\overline{h}_{\mathrm{T}}$ as a function of the area width $\ell/L_0$ for El Faro (black), Draupner (red) and Andrea (blue) sea states,  where $\overline{h}_{\mathrm{FST}}$ is the mean maximum surface height expected over the area $\ell^2$ during a sea state of duration $D=1$ hours and $\overline{h}_{\mathrm{T}}$ is the mean maximum surface height expected at a point. For comparisons, the empirical FST ratio from the El Faro HOS simulations (dashed line) together with the experimental observations at the Acqua Alta tower (squares) are also shown~\cite{fedele2013}. $L_0$ is the mean wavelength.

\item[Figure 10] HOS (squares) and theoretical (solid lines) predictions for the normalized nonlinear threshold $h_n/H_s$ exceeded with probability $1/n$; i) along the straight path $\Gamma$ spanned by El Faro while drifting at an estimated approximate average speed of~$2.5$~m/s over a time interval of 10 minutes (blue), ii) and also accounting for the vessel size~($241$~x~$30$~$m^2$)  (black), and over the strip area~($1500$~x~$30$~$m^2$) spanned by the vessel in a 10-minute interval (red). Confidence bands are also shown~(light dashes). Horizontal line denotes the threshold~$1.6H_s\approx14$~m, which is exceeded with probability~$3\cdot10^{-4}$,$1/400$ and~$1/100$ for the three cases shown.

\item[Figure 11] HOS simulations: expected spatial shape of a rogue wave whose crest height is~$>1.6 H_s\approx14$~m.

\end{description}

\end{document}